\documentstyle[12pt,epsf]{article}
\def\nm{\nonumber}
\def\llam{\lambda_1}
\def\lam{\Lambda_0}
\def\la{\Lambda_1}

\def\Del{\Delta}
\def\DDel{\Delta_1}

\def\beqa{\begin{eqnarray}}
\def\beq{\begin{equation}}
\def\F{\cal{F}}
\def\eeqa{\end{eqnarray}}
\def\eeq{\end{equation}}
\def\lab{\label}

\begin{document}

\begin{titlepage}
%\thispagestyle{plain}
%\pagenumbering{arabic}
\null
%\begin{flushright}
%March. 1996
%\end{flushright}
\vspace{2.0cm}
\begin{center}
{\Large \bf
Prepotential \\
of\\
$N=2$ $SU(2)$ Yang-Mills Gauge Theory\\
Coupled with a Massive Matter Multiplet}
\lineskip .80em
\vskip 5em
\normalsize
{\large Y\H uji Ohta}
\vskip 1.5em
{\it Department of Mathematics, Faculty of Science, Hiroshima University, \\
Higashi-Hiroshima 739, Japan}
\vskip 1.5em
\end{center}
\vskip3em

\begin{abstract}
We discuss the $N=2$ $SU(2)$ Yang-Mills theory coupled with a massive matter 
in the weak coupling. In particular, we obtain 
the instanton expansion of its prepotential. 
Instanton contributions in the mass-less limit are completely reproduced. We 
study also the double scaling limit of this massive theory and find that the 
prepotential with instanton corrections in the double scaling limit coincides 
with that of $N=2$ $SU(2)$ Yang-Mills theory without matter.
\end{abstract}
\end{titlepage}

\begin{center}
\section{Introduction}
\end{center}
\renewcommand{\theequation}{1.\arabic{equation}}\setcounter{equation}{0}

As is well-known, low energy properties of $N=2$ supersymmetric Yang-Mills 
gauge theory are dominated by a holomorphic function (prepotential) 
and were recently 
studied by Seiberg and Witten \cite{SW1,SW2}. In particular, they showed 
that the quantum moduli space was described by a kind of special geometry 
\cite{Fre} and identified the quantum moduli space of $N=2$ $SU(2)$ 
Yang-Mills theory without matter with the moduli space of a certain 
elliptic curve. Though they did not explicitly 
calculate the prepotential with instanton 
corrections, they qualitatively discussed the monopole and dyon 
masses, the metric on the quantum moduli space and a version of 
Olive-Montonen electric-magnetic duality and found that the strongly coupled 
vacuum turned out to be a weakly coupled theory of monopoles. 

After the discovery of \cite{SW1,SW2}, generalizations for another gauge 
groups 
\cite{PF,KLT,D,B,HO} and for $N=2$ $SU(2)$ Yang-Mills theory coupled with several 
mass-less matters \cite{IY} have been discussed and 
the instanton 
expansion of the prepotential has been found in \cite{KLT}, but discussions 
on the 
massive versions of \cite{KLT,IY} do not have been established quantitatively 
so far. For this reason, we can not say that the structure of quantum moduli 
space have been understood in detail even for the case of $N=2$ 
$SU(2)$ Yang-Mills theory 
when massive matters \footnote{We mean ``matters'' as the hypermultiplets of 
quarks in the fudamental representation of the gauge group.} 
are introduced. Thus we will study the quantum moduli spaces 
of $N=2$ $SU(2)$ Yang-Mills gauge theory coupled with several massive 
matters in this and subsequent papers \cite{O}. In particular, we will study 
the 
quantum moduli space of $N=2$ $SU(2)$ Yang-Mills theory coupled with a 
massive matter at weak coupling in this paper. 

The paper organizes as follows. In the next section, we derive the 
Picard-Fuchs equation for massive $N_f =1$ $N=2$ $SU(2)$ Yang-Mills theory
\footnote{$N_f$ is the flavour index.} and 
discuss the property of its solutions. It is noteworthy that the order of the 
differential equation is three in contrast with that of the 
mass-less theory whose order of 
the Picard-Fuchs equation is two. We also obtain the monodromy matrix near 
the weak coupling limit. In section 3, we derive 
the prepotential and its instanton expansion. The result coincides with 
\cite{IY} if the matter is mass-less. Considerations on double scaling limit 
of 
the $N_f =1$ massive theory are done in section 4. We will 
see that we can reproduce the 
instanton expansion of the $N_f =0$ theory. Last section 5 is summary.

\begin{center}
\section{$N_f =1$ Picard-Fuchs equation}
\end{center}
\renewcommand{\theequation}{2.\arabic{equation}}\setcounter{equation}{0}

Quantum moduli space of $N_f =1$ $N=2$ $SU(2)$ Yang-Mills theory can be 
described by 
a kind of hyperelliptic curve
	\beq
	y^2 =(x^2 -u)^2 -\la ^3 (x+m),\lab{curve}
	\eeq
and the meromorphic 1-form \footnote{
Note that normalization factor of our $\lambda_1$ 
is different from that of \cite{HO}.} which is given by 
	\beq
	\lambda_1 =\frac{\sqrt{2}xdx}
	{4\pi iy}\left[ \frac{x^2 -u}{2(x+m)}-2x \right], 
	\eeq
where $x,y \in \mbox{\boldmath$C$}$, $u$ is the gauge invariant parameter, 
$\la$ is a dynamical 
mass sacle of this theory and $m$ is the mass of the hypermultiplet \cite{HO}. 
Formulation by an elliptic curve can be found in \cite{SW2}.  

This curve has four branching points. 
In particular, in the weak coupling limit ($u \longrightarrow \infty$),
 they will be 
	\beqa
	x_1 &=& -\sqrt{u}-\frac{i\la^{3/2}}{2 u^{1/4}}+\frac{i\la^{3/2}m}
{4u^{3/4}} +\cdots ,\nm \\
	x_2 &=& -\sqrt{u}+\frac{i\la^{3/2}}{2 u^{1/4}}-\frac{i\la^{3/2}m}
{4u^{3/4}}+\cdots ,\nm \\
	x_3 &=& \sqrt{u}-\frac{\la^{3/2}}{2 u^{1/4}}-\frac{\la^{3/2}m}
{4u^{3/4}} +\cdots ,\nm \\
	x_4 &=& \sqrt{u}+\frac{\la^{3/2}}{2 u^{1/4}}+\frac{\la^{3/2}m}
{4u^{3/4}} +\cdots .
	\eeqa
Since we can take the cuts to run from $x_1$ to $x_2$ and $x_3$ to $x_4$, 
we may identify this 
curve as a genus one Riemann surface, as shown in the following figure 1.
	%\begin{figure}[h]
	%\begin{center}
	%\epsfile{file=tor.eps,scale=0.5}
	%\caption{The hyperelliptic curve as a genus one Riemann surface.}
	%\end{center}
	%\end{figure}
We then identify $\alpha$-cycle as a loop going around the cut from $x_4$ 
to $x_3$ counter 
clockwise and $\beta$-cycle from $x_3$ to $x_2$. 
As is obvious from the figure
, the intersection of these 
cycles is $\alpha \cap \beta =1$.
 
Now we can define periods $a(u),a_D (u)$ of $\lambda_1$ by 
	\beqa
	a(u) &=& \oint_{\alpha} \lambda_1 \label{a}, \\
	a_D (u) &=& \oint_{\beta} \lambda_1 \label{beta} .
	\eeqa
$a(u)$ is identified with the scalar component of the $N=1$ chiral multiplet 
and 
$a_D (u)$ is its dual. 
We are interested in their evaluation, but it is not so easy to accomplish it 
exactly. So we take a method of Picard-Fuchs equation. It is given by	
	\beqa
	& &\frac{d^3 \Pi_1}{du^3} +\frac{3\DDel(m) +\DDel'(m)(4m^2 -3u)}
	{\DDel(m) (4m^2 -3u)}
	\frac{d^2\Pi_1}{du^2}\nm \\
	&&-\frac{8\left[4(2m^2-3u)(4m^2-3u)+3(3\la^3 m-4u^2 )\right]}
	{\DDel(m)(4m^2-3u)}
	\frac{d\Pi_1}{du} =0 , \lab{pic1}
	\eeqa
where 
	\beq
	\DDel(m)=27\la^6 +256\la^3 m^3 -288\la^3 mu-256m^2u^2+256u^3 ,
	\eeq
and $\Pi_1 =\int_{\gamma} \lambda_1$, $\gamma$ is a suitable 1-cycle and 
$\DDel' =d\DDel/du$. 
Note that $\DDel (m)$ is the discriminant of the curve (\ref{curve}). 
It is easy to find that (\ref{pic1}) has no symmetry over the $u$-plane 
and the mass plays a role to break the symmetry. (\ref{pic1}) has obviously 
regular singular points which are solutions to $\Del_1 (m) =0$ and 
$4m^2 -3u=0$. These singular points correspond to mass-less states. 
Since we are going to treat only weak coupling limit in this paper, 
we do not discuss the 
behaviour of the moduli space near these 
singular points, but they should be discussed elsewhere. 
We have checked that (\ref{pic1}) can be also obtained as a result of the 
double scaling limit of the Picard-Fuchs equation of the massive 
$N_f =2$ theory. 

In the case of mass-less limit $(m \longrightarrow 0)$, 
this third order differential equation reduces to the second one,  
	\beq
	\frac{d^3 \Pi_1}{du^3} -\frac{\DDel(0) -\DDel'(0)u}{\DDel(0) u}
	\frac{d^2\Pi_1}{du^2}
	+\frac{64u}{\DDel(0)}\frac{d\Pi_1}{du} =0,
	\eeq
i.e,
	\beq
	(27\la^6 +256u^3 )\frac{d^2 \Pi_{1}}{du^2}+64u\Pi_1 =0 \lab{2.9},
	\eeq 		
where we set the integration constant as 0 because it can 
be shown directly. This equation has already been 
obtained in \cite{IY}. 

The mechanism of this reduction is explained as follows. When the matter is 
massive, 
$\lambda_1$ will acquire an extra simple pole corresponding to $x=-m$ in 
contrast with the mass-less 
case. At first sight, even if $m=0$, $\llam$ seems to have a pole at $x=0$, 
but the locus 
of the pole can be canceled out between denominator and numerator of $\llam$. 
Accordingly, in general, the number of poles 
of massive meromorphic 1-form is equal to that of the mass-less meromorphic 
1-form plus 1. 
Since the differentiation reduces the order by 1 \cite{Gri}, 
the reduction will require one step more when $\llam$ is massive. 
Therefore the order of the differential equation which periods of 
$\llam$ should satisfy will increase 
one more than that of mass-less $\llam$, i.e, the order will be
 three and this observation is consistent with 
(\ref{pic1}) and (\ref{2.9}). 

In order to get the solutions of (\ref{pic1}) near $u =\infty$, 
we take $z=1/u$. After this change of variable, we use Frobenius's method. 
Then we find that its indicial equation has three roots, i.e, $0,-1/2,-1/2$ 
(double roots). 
The solution $\rho_0 (z)$ corresponding to the index 0 is in fact trivial, 
i.e, it is 
a constant, 
	\beqa
	\rho_0 (z)= \epsilon \lab{trivial}.
	\eeqa
However, this constant $\epsilon$ may depend on $\la$ or $m$ and will be 
determined below. Geometrically, $\epsilon$ corresponds to the residue 
contribution of the pole of the meromorphic 1-form. Of course, $\epsilon$ 
must vanish in the mass-less limit 
because the mass-less theory does not have such a pole. 
Thus $\epsilon$ is a function of the mass. On the other hand, 
there are two independent solutions corresponding to the 
index $-1/2$. One of them is  
	\beqa
	\rho_1 (z)&=& z^{-1/2} \sum_{i=0}^{\infty}a_i z^i ,
	\eeqa
where the first several expansion coefficients $a_i$ are given 
in appendix A. 
We find that $a_n$ can be represented 
by a polynomial of $\la^{3i} m^j$ with 
$2n =3i+j$, where $i$ and $j$ are some non-negative integers. 
The other solution behaves logarithmic. It is
	\beqa
	\rho_2 (z) &=& \rho_1 (z) \ln z +z^{-1/2} \sum_{i=1}^{\infty}b_i z^i 
	\lab{r2}
	,\eeqa
where the first several coefficients $b_i$ are given in appendix B. 
Note that $b_n$ can also be represented by a polynomial of $\la^{3i}m^j$, 
where $i$ and $j$ are some non-negative integer with $2n =3i+j$. 
But in this time, $i$ and $j$ must move over all combinations. 

Since we would like to get $a(u)$ and $a_D (u)$, let us consider whether 
we can express them as linear combinations 
of $\rho_0 , \rho_1$ and $\rho_2$. 
First, in order to see an asymptotic behaviour of $a(u)$ near $u =\infty$, 
we must calculate the lower order expansion of 
the integral (\ref{a}). This is done in appendix C. 
Making a comparison $\rho_1 $ and $\rho_0$ 
with (\ref{exp}), we can see that 
	\beqa
	a(u) &=& n \rho_0 (z)+\frac{\rho_1 (z)}{\sqrt{2}} \lab{iden}
	,\eeqa
where we identified as
	\beq
	\rho_0 (z)=-\frac{\sqrt{2}}{4}m \lab{resi}
	.\eeq
It is easy to find that $a(u)$ can be expressed by a hypergeometric 
function in the mass-less limit \cite{IY}.  
$a_D (u)$ can be written as a linear combination 
of $\rho_0 ,\rho_1$ and $\rho_2$ by 
comparison it with (\ref{exp2}),
	\beqa
	a_D (u) &=& A\rho_2 (z)+B \rho_1 (z)+n' \rho_0 (z) , \lab{iden2}
	\eeqa
where
	\beqa
	A&=& -\frac{i3\sqrt{2}}{4\pi},\nm  \\
	B&=&  \frac{i\sqrt{2}}{4\pi}c \lab{AB}
	\eeqa 
and $c=-6 +8\ln 2 -i\pi -6\ln \la$. 

From these explicit expressions for the periods, we can easily find that 
the monodromy 
matrix near $u=\infty$ acts to the three objects as 
	\beq
	\left(\begin{array}{c}
	a_D  \\
	a \\
	\epsilon \end{array}\right) \longrightarrow \left( 
	\begin{array}{rrc}
	-1 & 3 & 2n' -3n \\	
	0& -1 & 2n \\
	0& 0& 1 
	\end{array}\right) \left(\begin{array}{c}
	a_D  \\
	a \\
	\epsilon \end{array}\right).
	\eeq
Note that the monodromy matrix is now quantized by the winding number $n$ 
and $n'$. Since these winding numbers are arbitrary, we may say 
that there are ``many'' monodromy matrices near 
$u=\infty$. This observation will be valied even for 
near the regular singular points.

\begin{center}
\section{Prepotential}
\end{center}
\renewcommand{\theequation}{3.\arabic{equation}}\setcounter{equation}{0}

Let us try to construct the prepotential ${\F}_1$ which is a solution 
to the following differential equation
	\beq
	a_D (u) =\frac{d {\F}_1}{d a}
	,\eeq
but we use a new variable $\tilde{a}=a-n\epsilon$ for convenience.  
For that purpose, first, we must express $u$ as a series of $\tilde{a}$. 
We can easily get it from (\ref{iden}), i.e, 
	\beqa
	u &=&2 \tilde{a}^2 + \frac{\la^3 m}{16 \tilde{a}^2}-\frac{3 \la^6}
{2048 \tilde{a}^4}  
	+\frac{5 \la^6 m^2}{4096 \tilde{a}^6}-\frac{7 \la^9 m}{65536 
\tilde{a}^8}
	+\frac{1}{\tilde{a}^{10}}\left(\frac{153 \la^{12}}{67108864} + 
	\frac{9\la^9 m^3}{131072}\right) \nm \\
	& &-\frac{715 \la^{12} m^2}
	{67108864 \tilde{a}^{12}}+\frac{1}{\tilde{a}^{14}}\left(\frac{1131 
\la^{15} m}
	{2147483648}+\frac{1469 \la^{12} m^4}{268435456}\right) \nm \\
	& & -\frac{1}{\tilde{a}^{16}}\left(\frac{1155 \la^{18}}{137438953472}+
	\frac{2625\la^{15} m^3}{2147483648}\right)\nm \\
	& &-\frac{1}{\tilde{a}^{18}}\left(\frac{667879 \la^{18} m^2}
{1099511627776}+
	\frac{148819 \la^{15} m^5}{17179869184}\right)+\cdots \lab{u}
	\eeqa
Inserting (\ref{u}) into (\ref{iden2}) and integrating it over $\tilde{a}$, 
we can obtain the prepotential
	\beq
	{\cal{F}}_1 =  
	\frac{i\tilde{a}^2}{\pi} \left[ \frac{3}{4}\ln \left(\frac
	{\tilde{a}}{\la}
	\right)^2 +\frac{3}{4}\left(-1+\frac{c}{3}+\ln 2 
	\right) -\frac{\sqrt{2}\pi}{4i\tilde{a}}n' m -\frac{m^2}{4\tilde{a}^2} 
	\ln \tilde{a} +
	\sum_{i=2}^{\infty} {\F}_{i}^1 \tilde{a}^{-2i} \right] \lab{pre}
	,\eeq
where the first several coefficients ${\F}_{i}^1$ are recorded in appendix D. 
Note that ${\F}_{n}^1$ is expressed again by a polynomial 
of $\la^{3i} m^j$ with $2n=3i+j$. 
We can find that our result (\ref{pre}) coincide with \cite{IY} in the 
mass-less limit. It is interesting to note that (\ref{pre}) has a curious 
term proportional to $(\ln \tilde{a}) /\tilde{a}^2$ in the brackets. 
However, we can find that \footnote{After the completion of this work, 
Prof.S.K.Yang pointed out this series expansion. } 
	\beqa
	{\F}_{s}^{1} &=& \left( \tilde{a} -\frac{m}{\sqrt{2}} \right)^2 
	\ln \left( \tilde{a} -\frac{m}{\sqrt{2}} \right) +
	\left( \tilde{a} +\frac{m}{\sqrt{2}} \right)^2 \ln 
	\left( \tilde{a} +\frac{m}{\sqrt{2}} \right) \nm \\
	& =&2\tilde{a}^2 \ln \tilde{a} +m^2 \ln \tilde{a} +
	\frac{3}{2}m^2 -\frac{m^4}{24 \tilde{a}^2} 
	-\frac{m^4}{240\tilde{a}^4}-
	\cdots \lab{poin}
	.\eeqa
Thus we may rewrite (\ref{pre}) as 
	\beqa
	{\cal{F}}_1 &=&  
	\frac{i\tilde{a}^2}{\pi} \left[ \frac{3}{4}\ln \left(\frac
	{\tilde{a}}{\la}
	\right)^2 +\frac{3}{4}\left(-1+\frac{c}{3}+\ln 2 
	\right) -\frac{\sqrt{2}\pi}{4i\tilde{a}}n' m \right.- 
	\frac{1}{4\tilde{a}^2}{\F}_{s}^{1}\nm \\
	& &+\frac{1}{2}\ln \tilde{a} +\left. \frac{3m^2}{8\tilde{a}^2} 
	+\sum_{i=2}^{\infty} \widetilde{\F}_{i}^1 \tilde{a}^{-2i} \right] 
	,\eeqa
where $\widetilde{\F}_{i}^1$ is $\la$ dependent part of ${\F}_{i}^1$.
\begin{center}
\section{Double scaling limit}
\end{center}
\renewcommand{\theequation}{4.\arabic{equation}}\setcounter{equation}{0}

In this section, we examine the double scaling limit of the massive $N_f =1$ 
theory discussed in previous sections. 

To begin with, let us discuss the Picard-Fuchs equation. 
Since the $N_f =1$ curve turns to the $N_f =0$ 
curve in the double scaling limit ($ m \longrightarrow \infty,\la 
\longrightarrow 0, m\la^3 =\Lambda_{0}^4 $ fixed, where $\Lambda_0$ is a 
dynamical parameter 
of the $N_f =0$ theory), in other words,
 $N=2$ $SU(2)$ Yang-Mills theory is considered as a low energy theory 
of the massive $N_f =1$ theory \cite{SW2}, we may expect that 
(\ref{pic1}) reduces to the Picard-Fuchs equation of the $N_f =0$ 
theory. In fact, we find that (\ref{pic1}) in the double scaling 
limit is given by 
	\beq
	\frac{d^3 \Pi_1}{du^3} 
	+\frac{3\cdot 256m^2(\lam^4 -u^2)-2\cdot 256m^2 u\cdot 
 	4m^2}{256m^2 (\lam^4 -u^2)\cdot 4m^2}\frac{d^2\Pi_1}{du^2} 
	-\frac{32\cdot 8m^4}{256m^2 (\lam^4  -u^2)\cdot 4m^2}
	\frac{d\Pi_1}{du} =0,
	\eeq
i.e,
	\beq
	 \frac{d^2 \Pi_{1}}{du^2}-\frac{1}{4(\lam^4 -u^2)}\Pi_1  =\mbox
{constant} \lab{red}.
	\eeq
(\ref{red}) shows the global $\mbox{\boldmath$Z$}_2$ symmetry over the 
$u$-plane. 
At first sight, this can be seen as the 
Picard-Fuchs equation of $N=2$ $SU(2)$ Yang-Mills theory without 
matter \cite{KLT,EY}. 
However, we can not say that $\Pi_1$ also reduces 
to that of $N_f =0$ Picard-Fuchs equation because 
there is no reason why the relation 
	\beqa
	\Pi_1 (u,m,\la)\stackrel{\rm double\  scaling \ limit}{-------
	\rightarrow} 
	\Pi_0 (u,\Lambda_0 ) \lab{(6)}
	\eeqa
should hold, where $\Pi_0$ is a period integral of 
$N_f =0$ theory. So we can not insist on that the integration 
constant in the right hand side of (\ref{red}) must be 0. 

More precisely speaking, the meromorphic 1-form $\llam$ in the double 
scaling limit 
will behave as  
	\beqa
	\llam &\longrightarrow & \frac{\sqrt{2}xdx}{4\pi i\tilde{y}}
	\left[ \frac{x^2 -u}{2m}\left( 1-\frac{x}{m}+\frac{x^2}{m^2}-\cdots 
	\right) -2x \right] \nm \\
	& =& \frac{\sqrt{2}x(x^2 -u)dx}
	{8\pi im \tilde{y}}\left( 1-\frac{x}{m}+\frac{x^2}{m^2}-\cdots 
	\right)-\frac
	{\sqrt{2}x^2 dx}{2\pi i \tilde{y}}
	\lab{Nf}	
	,\eeqa
where $\tilde{y}^2 =(x^2 -u)^2 -\Lambda_{0}^4$ is the curve for 
the $N_f =0$ theory. 
The first term in the last expression 
is an ``extra'' 1-form which depends on the mass $m$ 
while the second is nothing other than the meromorphic 1-form of 
the $N_f =0$ theory. Therefore, 
naively speaking, the solutions to (\ref{pic1}) 
consist of the contributions originated from this extra 1-form 
and $\lambda_0$, in the double scaling limit. In fact, (\ref{trivial}) 
and (\ref{r2}) 
diverges to infinity in the double scaling limit as is easy to find. 
Accordingly, when we discuss the low energy limit of $N_f =1$ theory, 
we must carefully treat (\ref{(6)}). 
Though $\rho_0 ,\rho_1 $ and $\rho_2$ are indeed solutions to (\ref{pic1}) 
for finite $\la$ and $m$, nothing gives an assurance 
for that they constitute fundamental solutions even for infinitely 
large $\la$ or $m$. Recall that we have obtained the 
solutions to (\ref{pic1}), assuming that $\la$ and $m$ are finite. 
This fact will be reflected on the solutions to 
the Picard-Fuchs equation. Consequently, there must be a gap between $\Pi_1$ 
in the double scaling limit and $\Pi_0$. This gap will appear as 
divergence due to the mass.  

Next, let us directly examine the above observations, focusing on the 
periods $a(u)$ and $a_D (u)$. As for the residue part of the 
periods, i.e, $\rho_0$, it moves to infinity due to $m \longrightarrow 
\infty$. However, as is discussed in \cite{SW2}, the quark with large mass 
can be integrated out, so we can 
eliminate the residue dependence of the periods. We can see that among the 
expansion coefficients $a_n $ $(n>1)$ of $a(u)$, only the coefficients 
of even degree survive in the double scaling 
limit and those of the odd degrees vanish. Accordingly, 
we may then say that $a(u)$ converges to a solution to (\ref{red}). 
In other words, $a(u)$ is not affected by the contributions from the extra 
1-form. This fact suggests that $a(u)$ has a nice property which is valid 
under the double scaling. It is easy 
to check that the period $a(u)$ in the double scaling limit can be again 
expressed as a hypergeometric function. 
In order to see the behaviour of $a_D (u)$ in the double scaling limit, 
we must rewrite it as a series of $a(u)$ and then take the double scaling 
limit. We can easily see that the expansion coefficients $b_n$'s diverge to 
infinity in the double 
scaling limit. This means that the contributions for $a_D (u)$ from 
the extra 1-form are non-trivial. 
Since it is hard to see the difference of it 
with the period over $\beta$-cycle of the $N_f =0$ theory in this 
situation, we should first arrange $a_D (u)$ with some ``good'' variable. 
For that purpose, we can take $a(u)$ as the good variable. From 
(\ref{u}), $a_D (u)$ will be expanded as 
	\beqa
	a_D (u) &=& n' \epsilon + 
	\sqrt{2} a [B-A\ln 2 -2A\ln a]+\frac{A}{\sqrt{2}} \left[ \frac{m^2}
{3a} +\frac{m}{72a^3}(-3\la^3 +2m^3 )\right. \nm \\
	& &\left. +\frac{1}{46080a^5}(45\la^6 +256m^6 )
	+\frac{m^2}{86016a^7}(-105\la^6 +128m^6 ) +\cdots \right] ,\lab{expand}
	\eeqa
where $\epsilon,A$ and $B$ are given in (\ref{resi}) and (\ref{AB}). 
Note that each coefficient of $a^{-2i+1}$ $(i>0)$ consists of a finite 
part and a ``divergent'' 
part and the latter is always proportional to $m^{2l},\ l \in 
\mbox{\boldmath$N$}$. 
However, since 
the heavy quark with large mass must be 
integrated out \cite{SW2} as have been mentioned above, 
it would be enough to consider the finite parts. 
The divergence due to the large mass can be eliminated in that sence. 
Accordingly, if we extract only finite contributions 
we can arrive at a correct answer to get the periods $a_D (u)$ 
of the $N_f =0$ theory. 
The reader may ask how to deal with the constants $A$ and $B$ in 
(\ref{expand}) under the double scaling. 
These constants should be replaced with that of the $N_f =0$ theory. 
This is because 
the initial conditions for 
the Picard-Fuchs equation of the $N_f =1$ theory are 
different from that of the $N_f =0$ theory. 
In this way, $\Pi_1$ reduces to $\Pi_0$ and under this situation 
the integration constant in (\ref{red}) will be 0. Then (\ref{red}) 
is nothing other than the Picard-Fuchs equation of the $N_f =0$ theory. 
Of course, in this time we must change $\Pi_1$ to $\Pi_0$ in (\ref{red}). 

Let us examine the double scaling limit of the prepotential. 
The procedure to do it consists of two steps. The first one is 
to take the double scaling limit 
of the first four terms in the brackets in (\ref{pre}). The second one is to 
consider the double scaling limit of the coefficients 
of $a^{-2i}$. 

In order to accomplish the first step, we use a trick. Recall that we can 
add or minus 
infinity related to the mass because of the reason described before. Thus, 
	\beqa
	\frac{3}{4}\ln \left(\frac{a}{\la}\right)^2 +\frac{3}{4}C &= &
	\frac{3}{4}\ln \left(\frac{a}{\la}\right)^2 
	+\frac{3}{4}C +\ln \left[\left(\frac{a}{m}
	\right)^{1/2} \cdot e^{D/2 -3C/4}\right] \nm \\
	 & =&\ln \left(\frac{a}{\Lambda_0} \right)^2 +\frac{D}{2}
	\eeqa
where $C=-1+\ln 2 +c/3$, $D=-6+6\ln2$ \footnote{The number ``$D$'' is 
different from that of \cite{KLT}, but it seems to be their mistake although 
it is not so important.} and $\tilde{a}$ is now replaced with $a$. 
Note that this trick essentially 
corresponds to the replacement of the ``initial'' 
conditions described 
above. Since the remaining two terms can be integrated out, we have dropped 
them here.   

For the expansion coefficients $\widetilde{\F}_{i}^1$, 
we can easily find that they will then be, 
	\beqa
	\widetilde{\F}_{2}^1 &\longrightarrow & -\frac{\Lambda_{0}^4}{64} \nm,
	 \\
	\widetilde{\F}_{3}^1 &\longrightarrow & 0,\nm \\
	\widetilde{\F}_{4}^1 &\longrightarrow & -\frac{5\Lambda_{0}^8}{32768}
	,\nm \\
	\widetilde{\F}_{5}^1 &\longrightarrow & 0,\nm \\
	\widetilde{\F}_{6}^1 &\longrightarrow & 
	-\frac{3\Lambda_{0}^{12}}{524288} ,\nm \\
	\widetilde{\F}_{7}^1 & \longrightarrow & 0, \nm \\
	\widetilde{\F}_{8}^1 &\longrightarrow  & -\frac{1469\Lambda_{0}^{16}}
	{4294967296}, \nm \\
	\widetilde{\F}_{9}^1 &\longrightarrow &0
	.\eeqa
Then we can get the following 
``renormalized'' prepotential $\widehat{\F}_1$ in the double scaling limit
	\beq
	\widehat{{\cal{F}}}_{1}=\frac{ia^2}{\pi} \left[\ln \left(
	\frac{a}{\Lambda_0}\right)^2 
	+\frac{D}{2}-\frac{\Lambda_{0}^4}{2^6 \cdot a^4}-
	\frac{5\Lambda_{0}^8}{2^9 \cdot 
	a^6}-\frac{3\Lambda_{0}^{12}}{2^{19}\cdot a^8}-
	\frac{1469\Lambda_{0}^{16}}
	{2^{32} \cdot a^{10}}-\cdots \right] 
	.\eeq
This agrees with the result of \cite{KLT}.

\begin{center}
\section{Summary}
\end{center}
\renewcommand{\theequation}{5.\arabic{equation}}\setcounter{equation}{0}

We have studied the moduli space of 
$N=2$ $SU(2)$ Yang-Mills theory coupled with a matter 
multiplet at weak coupling. In particular, we have determined its 
prepotential and monodromy matrix. For general values of $\la$ and $m$, 
we have established that the two periods of the meromorphic 1-form 
can be written as 
	\beqa
	a(u) &=& -\frac{\sqrt{2}}{4}nm 
+\frac{1}{2}\sqrt{2u}\left[ 1+\sum_{i=2}^{\infty} a_i (\la^3 ,
	m) u^{-i} \right] ,\nm \\
	a_D (u) &=& -\frac{\sqrt{2}}{4}n' m +
	\frac{3i}{2\pi}\tilde{a}(u) \ln \left(\frac{u}{\la^2}\right)
	+\sqrt{u} \sum_{i=0}^{\infty} a_{D_i} (\la^3 ,m) u^{-i} ,
	\eeqa
where $a_i (\la^3 ,m)$ and $a_{D_i} (\la^3 ,m)$ are homogeneous 
polynomials of order $2i$, 
instead of the formulae noted in \cite{SW2,IY}. And we have proposed 
the exact expression for 
the prepotential as in (\ref{pre}). This prepotential has a 
curious term, as we have already seen, so it should be examined by 
some field theoretical method. 
The coefficients of instanton expansion in the mass-less limit 
completely coincide with \cite{IY}. On the other hand, we have succeeded in 
constructing the $N_f =0$ theory as a low energy theory 
of the massive $N_f =1$ theory and have found 
that we can recover the instanton expansion of the prepotential 
of the $N_f =0$ theory.

Finally, we give some comments. 
Since the massive $N_f =1$ theory can be considered as a low energy theory of 
the massive $N_f =2$ theory \cite{SW2}, 
all our results will be expected to be reproduced from it. In addition to 
this, it will be interesting to reconstruct our results in the languages 
of integrable systems such as Whitham hierarchy and so on \cite{EY,IM}. The 
discussions in this paper should be compared with those approaches, 
but such considerations unfortunately are not proceeded at present.  
 
\begin{center}
\section*{Acknowledgement}
\end{center}

The author would like to give his thanks to Dr.H.Kanno for helpful comments 
and suggestions.

\begin{center}
\section*{Appendix A Expansion coefficients $a_i$}
\end{center}
\renewcommand{\theequation}{A\arabic{equation}}\setcounter{equation}{0}

The first several coefficients of $\rho_1$ are
	\beqa
	a_0 &=& 1 ,\nm \\
	a_1 &=& 0 ,\nm \\
	a_2 &=& -\frac{\la^3 m }{16} ,\nm \\
	a_3 &=& \frac{3 \la^6 }{1024},\nm \\ 
	a_4 &=& -\frac{15\la^6 m^2  }{1024},\nm \\
	a_5 &=& \frac{35 \la^9 m}{16384} ,\nm \\
	a_6 &=& -\frac{105\la^9}{4194304}(3\la^3 +256 m^3 ),\nm \\
	a_7 &=& \frac{3465\la^{12} m^2 }{2097152} ,\nm \\
	a_8 &=& -\frac{3003\la^{12} m}{67108864}(3\la^3 +80m^3 ) ,\nm \\
	a_9 &=& \frac{15015\la^{15}}{4294967296}(\la^3 +384m^3 )
	.\eeqa 

\begin{center}
\section*{Appendix B Expansion coefficients $b_i$ }
\end{center}
\renewcommand{\theequation}{B\arabic{equation}}\setcounter{equation}{0}

The first several coefficients of $\rho_2$ are 	
	\beqa
	b_1 &=& \frac{1}{3}m^2  ,\nm \\
	b_2 &=& \frac{m}{72}(3\la^3 +4m^3 ),\nm \\
	b_3 &=& \frac{1}{23040}(-45\la^6 +480\la^3 m^3 +512m^6 ), \nm \\
	b_4 &=& \frac{m^2 }{21504}(-21\la^6 +224\la^3 m^3 +256m^6 ),\nm \\
	b_5 &=& \frac{m}{276480}(-120\la^9 +1575\la^6 m^3 
	+1920\la^3 m^6 +2048m^9 ),\nm \\
	b_6 &=& \frac{1}{415236096}(9801\la^{12} -937728\la^9 m^3 
	+1419264\la^6 m^6 \nm \\
	& &+2162688\la^3 m^9 +2097152m^{12}),\nm \\
	b_7 &=& \frac{m^2}{5725224960}(323505\la^{12} +13453440\la^9 m^3 
	+15375360\la^6 m^6 \nm \\
	& & +23855104\la^3 m^9 +20971520m^{12}) ,\nm \\
	b_8 &=& \frac{m}{3019898880}(45837\la^{15}-5636520\la^{12} m^3 
	+4392960\la^9 m^6 \nm \\
	& &+7028736\la^6 m^9 +10485760\la^3 m^{12}  +8388608m^{15} )
	,\nm \\
	b_9 &=& \frac{1}{13799729922048}(-9689337\la^{18} +3483611712\la^{15}
	m^3 \nm \\
	& & +16467010560\la^{12}m^6 +16728391680\la^9 m^{9} +
	29198647296\la^6 m^{12} \nm \\
	& &+41070624768 \la^3 m^{15} +30064771072m^{18} ).
	\eeqa
 
\begin{center}
\section*{Appendix C Lower order expansion of the periods}
\end{center}
\renewcommand{\theequation}{C\arabic{equation}}\setcounter{equation}{0}

In this appendix, we show that the lower order expansion of the period 
integral (\ref{a}) in detail 
as an example. However, we must treat (\ref{a}) carefully 
because this $\alpha$-cycle is defined to be an usual homology basis. 
Recall that the meromorphic 1-form was constructed under the assumption 
such that the asymptotic behaviour of $a(u)$ 
at $u=\infty$ was 
to be $a(u) \sim \sqrt{u/2}$ even if the theory was massive \cite{HO}. 
Therefore even if we evaluate the period (\ref{a}) by direct calculation, 
we can not obtain the correct contribution from the pole. 
When the cycle may across the pole, then the integration 
must pick up the residue of the pole. However, since the $\alpha$-cycle 
in (\ref{a}), as we have stated above, avoids the pole, we must study 
the case such that the cycle deforms from $\alpha$ to $\alpha'$ which 
enclose the pole and the two branching points $x_3$ and $x_4$ as 
shown in the following figure 2.
 	%\begin{figure}[h]
	%\begin{center}
	%\epsfile{file=cycle.eps,scale=0.5}
	%\caption{Deformations of the $\alpha$-cycle on the $x$-plane.}
	%\end{center}
	%\end{figure}
As is easily seen from this figure, the direction of $\alpha'$ is the same 
as that of $\delta$ which enclose only the pole. 
However, taking into account of an effect for 
topological deformation, we can find that the $\alpha$-cycle 
can be identified with the loop $\alpha''$ in the figure. Namely, if 
the $\alpha$-cycle should 
move on the another covering of this $x$-plane and back onto the original 
one, it will enclose $x_1$ and $x_2$, i.e, another cut. 
But in this time, the directions of 
$\alpha$ and $\alpha''$ will be different. Therefore when $\alpha''$ 
across the pole, the direction of $\alpha'''$ which is 
a deformation of $\alpha''$ 
and that of $\delta$ will be different. 
This fact causes to change 
the sign of the residue. The reader may ask that 
the sign of the integral over the $\alpha''$ should be reflected. 
Of course it is right. But since we usually use the convention 
such that the overall sign of $a(u)$ without the residue contribution, 
for example, to be $+$, e.g, 
$a(u) \sim +\sqrt{u/2}$ in $N=2$ $SU(2)$ Yang-Mills theory, 
we should change the sign of the residue instead of that of 
the integral in order to preserve the convention. 

From these discussions, the true expression for the massive period $a(u)$ 
should be defined by
	\beqa 
	a(u) &:=& \oint_{\tilde{\alpha}} \lambda_1 \nm \\
	&=&\oint_{\alpha} \lambda_1 +\oint_{\delta} \lambda_1 \nm \\
	&=&\oint_{\alpha} \lambda_1 +2\pi i n\cdot \mbox{Res}\left. 
	(\lambda_1 )\right|_{\mbox{\scriptsize at} \ x=-m} 
	\lab{true},
	\eeqa
where $\tilde{\alpha}$ is a certain member of the family of $\alpha$-cycle 
which may include the pole and the cut 
inside the loop, $\delta$ is a small loop around the pole and 
$n=0,\pm 1$. If $\tilde{\alpha}$ avoids the 
pole, then $\tilde{\alpha} =\alpha$ and $n=0$. 
If $\tilde{\alpha}$ enclose the pole and the directions 
of $\tilde{\alpha}$ and $\delta$ coincide, $n=+1$. If the directions are 
different while $\tilde{\alpha}$ enclose the pole, then $n=-1$. 
To clarify, we should further comment on the number $n$. 
We have treated only the case such that $\tilde{\alpha}$ winds once 
around the pole, but we may also allow the case such that 
it winds several times around the pole. In this time, 
$n$ can be interpreted as winding number and will be 
$n \in \mbox{\boldmath$Z$}$.  

Let us evaluate (\ref{true}). First, note that 
	\beq
	\oint_{\alpha} \llam 
	= 2 \int_{x_4}^{x_3} \llam  \lab{int} .
	\eeq
In the right hand side, the factor 2 is required because the integral 
over $\alpha$-cycle contains an integral from $x_4$ to $x_3$ and from $x_3$ to 
$x_4$ on the other side of the cut. In order to calculate (\ref{int}), 
we introduce a new variable $t$ such as $x=\sqrt{u} t$. Then (\ref{int}) 
will be 
	\beqa
	\oint_{\alpha} \lambda_1 
	&=& \frac{2\sqrt{2}}{4\pi i}\int_{x_4 /\sqrt{u}}^{x_3 /\sqrt{u}} 
	\frac{utdt}{\sqrt{u^2 (t^2 -1)^2 -\la^3 (\sqrt{u}t+m)}} 
	\left[ \frac{u(t^2 -1)}{2(\sqrt{u}t +m)} -2\sqrt{u}t \right] \nm \\
	&=& \frac{i}{4\pi}\sqrt{2u} \int_{x_4 /\sqrt{u}}^{x_3 /\sqrt{u}} 
	\frac{(3\sqrt{u}t^3 +4mt^2 +\sqrt{u}t )}{\sqrt{u}t +m} \nm \\
	& &\times 
	\left[ \frac{1}{t^2 -1}+\frac{\la^3 m}{2u^2 (t^2 -1)^3}+\frac{3\la^6 
	t^2}{8u^3 (t^2 -1)^5} + \cdots \right]dt \nm \\
	&=& \frac{1}{2}\sqrt{2u} \left( 1-\frac{\Lambda_{1}^3 m}{16u^2}+
	\cdots 
	\right). 
	\eeqa 	
Taking the contribution from the pole into account, 
we can arrive at 
	\beq
	a(u)= -\frac{\sqrt{2}}{4}nm +\frac{1}{2}\sqrt{2u} 
	\left( 1-\frac{\Lambda_{1}^3 m}{16u^2}+
	\cdots 
	\right)\lab{exp}.
	\eeq

On the other hand, the integration over $\beta$-cycle is not well-defined, 
as well. 
This can be seen by evaluating its lower order expansion. (\ref{beta}) will be 
	\beq
	\oint_{\beta} \llam =	\frac{i}{4\pi}\sqrt{2u}\left( 3\ln u +
	8\ln 2 -i\pi -6 -6\ln \la +\frac{im\pi}{\sqrt{u}} +\cdots \right)
	\lab{wrong} .\eeq
At first sight, this integration seems to be 
$a_D (u)$ with the contribution from the pole. In fact, this observation 
is not wrong. 
However, (\ref{wrong}) does not contain the possibilities 
such that the $\beta$-cycle does not enclose the pole, for example. 
In other words, 
the topological deformation of $\beta$-cycle as in the case of 
$\alpha$-cycle is (partially) ignored. Since the pole 
merely contributes as only constant term, (\ref{wrong}) will be 
well-defined as the 
period over the $\beta$-cycle avoiding the pole, 
if the constant term is extracted. Therefore the true definition of 
$a_D (u)$ will be 
	\beqa
	a_D (u) &:= &\oint_{\tilde{\beta}} \llam \nm \\
	&=& \oint_{\beta} \llam +\oint_{\delta} \llam \nm \\
	&=& \oint_{\beta} \llam  +2\pi i n' \cdot \mbox{Res}\left.(\llam)
	\right|_{\mbox{\scriptsize at} \ x=-m},\lab{last}
	\eeqa
where $\tilde{\beta}$ is a certain member of $\beta$-cycle 
which may enclose the pole and 
the cut, $\beta$ in (\ref{last}) means now a loop avoiding the pole 
and $n' =0,\pm 1$. If the loop $\tilde{\beta}$ winds 
around the pole several times, then $n' \in 
\mbox{\boldmath$Z$}$. In this way we can arrive at  
	\beq
	a_D (u) =-\frac{\sqrt{2}}{4}n' m +\frac{i}{4\pi}\sqrt{2u}\left( 
	3\ln u +
	8\ln 2 -i\pi -6 -6\ln \la  +\cdots \right)
	.\lab{exp2}
	\eeq

\begin{center}
\section*{Appendix D Expansion coefficients ${\F}_{i}^1$ }
\end{center}
\renewcommand{\theequation}{D\arabic{equation}}\setcounter{equation}{0}

First several coefficients of the prepotential (\ref{pre}) are listed below.
	\beqa
	{\F}_{2}^1 &= & -\frac{1}{64}\la^3 m +\frac{m^4}{96} ,\nm \\
	{\F}_{3}^1 &=&\frac{3\la^6}{16384}+\frac{m^6}{960} , \nm \\
	{\F}_{4}^1 &= & -\frac{5\la^6 m^2}{32768}+\frac{m^8}{5376} ,\nm\\
	{\F}_{5}^1 &=&\frac{7\la^9 m}{786432}+\frac{m^{10}}{23040} , \nm \\
	{\F}_{6}^1 &= &-\frac{153\la^{12}}{1073741824} -\frac{3\la^9 m^3}{
	524288} +\frac{m^{12}}{84480} ,\nm \\
	{\F}_{7}^1 &= &\frac{715\la^{12}m^2}{1073741824}+\frac{m^{14}}{279552}
	, \nm \\
	{\F}_{8}^1 &= & -\frac{1131\la^{15}m}{42949672960}-
	\frac{1469\la^{12}m^4}{4294967296}+\frac{m^{16}}{860160}, \nm \\
	{\F}_{9}^1 &= &\frac{385\la^{18}}{1099511627776}+\frac{525\la^{15}m^3}{
	8589934592}+\frac{m^{18}}{2506752}.  
	\eeqa 
	
\begin{center}

\end{center}
\end{document}